\documentclass[12pt,number,sort&compress]{elsarticle-published}
\biboptions{sort&compress}

\usepackage[psamsfonts]{amssymb}

\usepackage{amssymb}

\usepackage{bm}

\usepackage{siunitx}

\journal{Nucl. Instr. and Meth. A}

\usepackage{hyperref}
\usepackage{comment}

\newcommand{\gsim}{\hbox{ \raise3pt\hbox to 0pt{$>$}\raise-3pt\hbox{$\sim$} }}
\newcommand{\lsim}{\hbox{ \raise3pt\hbox to 0pt{$<$}\raise-3pt\hbox{$\sim$} }}
\newcommand{\del}{\ifmmode{\nabla}         \else{$\nabla$ }               \fi}

\newcommand{\figdir}{./}
\usepackage{graphicx}

\begin{document}

\begin{frontmatter}

\title{Measurement of the electron transmission rate of the gating foil \\
for the TPC of the ILC experiment}

\begin{frontmatter}
\title{Measurement of the electron transmission rate of the gating foil \\
for the TPC of the ILC experiment}
\end{comment}

%
%***************
%   Authors
%***************
%
\author[3]{M.~Kobayashi\corref{cor1}}
     \ead{makoto.kobayashi.exp@kek.jp}
     \cortext[cor1]{Corresponding author.
                           Tel.: +81 29 864 5376; fax: +81 29 864 2580.}
\author[7]{T.~Ogawa}
\author[11]{A.~Shoji}
\author[7]{Y.~Aoki}
\author[1,13]{K.~Ikematsu}
\author[1,14]{P.~Gros}
\author[8]{T.~Kawaguchi}
\author[12]{D.~Arai}
\author[12]{M.~Iwamura}
\author[12]{K.~Katsuki}
\author[12]{A.~Koto}
\author[12]{M.~Yoshikai}
\author[3]{K.~Fujii}
\author[1]{T.~Fusayasu}
\author[4]{Y.~Kato}
\author[10]{S.~Kawada}
\author[3]{T.~Matsuda}
\author[11]{S.~Narita}
\author[11]{K.~Negishi}
\author[17]{H.~Qi}
\author[5]{R.D.~Settles}
\author[1]{A.~Sugiyama}
\author[8]{T.~Takahashi}
\author[3,15]{J.~Tian}
\author[6]{T.~Watanabe}
\author[9]{R.~Yonamine}
%
%***************
%   Addresses
%***************
%
\address[3]{High Energy Accelerator Research Organization (KEK), Tsukuba 305-0801, Japan}
\address[7]{The Graduate University for Advanced Studies (Sokendai), Tsukuba 305-0801, Japan}
\address[11]{Iwate University, Iwate 020-8551, Japan}
\address[1]{Saga University, Saga 840-8502, Japan}
\address[8]{Hiroshima University, Higashi-Hiroshima 739-8530, Japan}
\address[12]{Fujikura Ltd., 1440, Mutsuzaki, Sakura-city, Chiba 285-8550, Japan}
\address[4]{Kindai University, Higashi-Osaka 577-8502, Japan}
\address[10]{Deutsches Elektronen-Synchrotron (DESY), D-22607 Hamburg, Germany}
\address[17]{Institute of High Energy Physics, Chinese Academy of Sciences,
Beijing 100049, China}
\address[5]{Max Planck Institute for Physics, DE-80805 Munich, Germany}
\address[6]{Kogakuin University, Shinjuku 163-8677, Japan}
\address[9]{Department of Physics, Tohoku University, Sendai 980-8578, Japan}
\address[13]{now at Institute of Multidisciplinary Research for Advanced Materials (IMRAM), Tohoku University, Sendai 980-8577, Japan}
\address[14]{now at Department of Physics, Engineering Physics \& Astronomy,
                    Queen's University, Kingston, Ontario K7L 3N6, Canada}
\address[15]{now at ICEPP, University of Tokyo, Hongo, Tokyo 113-0033, Japan}

\appendix
\setcounter{figure}{0}

\section{Estimation of Fano factor and avalanche fluctuation}

In the test of the gating foil using the $^{55}$Fe source, the electron
transmission rate is determined from the ratio of the photo-peak positions,
one for the photons converted upstream of the gating foil (gated signals)
and the other for those converted downstream (ungated signals). 
The energy resolution for the peak corresponding to the former photons gets worse
with decreasing transmission rate.

In addition to the gating or switching function,
the gating foil is capable of controlling the average fraction of 
signal electrons to be detected after gas amplification.  
In this appendix the energy resolution obtained with the gating foil is
analytically evaluated in terms of the Fano factor \cite{Fano},
the transmission rate of the gating foil, and the avalanche fluctuation in the GEM stack.
The value of the Fano factor and the size of the avalanche
fluctuation are then estimated from the observed resolution degradation with 
decreasing transmission rate\footnote{
We found that a similar approach had already been proposed by
M.A.~Chefdeville
for the measurement of avalanche fluctuations
with Micromegas, assuming a certain value for $F$ \cite{Chefdeville}.
}.

Let $n_0$ be the number of electrons liberated by
the conversion of an X-ray photon (Mn K$_\alpha$ or K$_\beta$),
and $p$ be the average transmission rate of the gating foil.
In fact, $n_0$ fluctuates around its average $\left< n_0 \right>$.
Then the average and the variance of the number of electrons passing
through the gating foil ($n$) are given by
\begin{eqnarray}
\left< n \right> &=& p \cdot \left< n_0 \right> \\
\sigma_n^2 &=& \left< \left( n - p \cdot \left< n_0 \right> \right)^2 \right> \\
  &=& \left< \left( n - p\cdot n_0 + 
         p \cdot \left( n_0 - \left< n_0 \right> \right) \right)^2 \right> \\
  &=& \left< n_0 \right> \cdot p \cdot \left( 1 - p \right) 
             + p^2 \cdot \sigma_{n_0}^2 \;.
\end{eqnarray}
In Eq.~(A.4), a binomial distribution is assumed for the transmission
through the gating foil\footnote{
Actually, the transmission rate depends on the photon's conversion point
relative to the hole center of the gating foil.
The measured value is expected to be the average transmission
since the conversion points of the uncollimated photons
spread wide enough
as compared to the hole pitch, and the diffusion of created electron is
comparable to the size of the holes in most cases.
}, 
and the average was taken over $n_0$ after averaging over $n$.  

The surviving electrons are then gas-amplified by the GEM stack and
detected by the readout pads.
The induced charge on the pads ($Q$) is given by
\begin{equation}
Q = \sum_{i=1}^n q_i + \delta_q
\end{equation}
where $q_i$ is the gas-amplified charge of the $i$-th drift electron
and $\delta_q$ denotes the random electronic noise charge on the pads
with $\left< \delta_q \right> = 0$, which is 
uncorrelated with the signal charge.
Assuming the same gas gain and its fluctuation for all the drift electrons,
the average and the variance of the induced charge are given by
\begin{eqnarray}
\left< Q \right> &=& \left< n \right> \cdot \left< q \right> \\
  &=& p \cdot \left< n_0 \right> \cdot \left< q \right> \\
\sigma_Q^2 &=& \left< \left( \sum_{i=1}^n q_i + \delta_q 
             - \left< n \right> \cdot \left< q \right> \right)^2 \right> \\
  &=& \left< \left( \sum_{i=1}^n \left( q_i - \left< q \right> \right)
       + \left( n - \left< n \right> \right) \cdot \left< q \right> 
       + \delta_q  \right)^2 \right> \\
  &=& \left< n \right> \cdot \sigma_q^2 + \left< q \right>^2 \cdot \sigma_n^2
      + \left< \delta_q^2 \right> \\
  &=& p \cdot \left< n_0 \right> \cdot \sigma_q^2
       + \left< q \right>^2 \cdot \left< n_0 \right> \cdot p \cdot \left( 1 - p \right)
       + \left< q \right>^2 \cdot p^2 \cdot \sigma_{n_0}^2 + \sigma_{\delta_q}^2 \;. 
\end{eqnarray}

Accordingly, the energy resolution squared ($R^2\;$) is given by
\begin{eqnarray}
R^2 &\equiv& \frac{\sigma_Q^2}{\left< Q \right>^2} \\
 &=& \frac{1}{p \cdot \left< n_0 \right>} \cdot \frac{\sigma_q^2}{\left< q \right>^2}
   \; + \; \frac{\left( 1 - p \right)}{p \cdot \left< n_0 \right>}
   \; + \; \frac{\sigma_{n_0}^2}{\left< n_0 \right>^2}
   \; + \; \frac{\sigma_{\delta_q}^2}{p^2 \cdot \left< n_0 \right>^2 \cdot 
               \left< q \right >^2} \\
  &=& \frac{1}{\left< n_0 \right>} \cdot \left( F - 1 \; + \; \frac{1+f}{p} \right)
   \; + \; \frac{\sigma_{\delta_q}^2}{p^2 \cdot \left< n_0 \right>^2 \cdot 
               \left< q \right>^2} \\
  &=& \frac{F-1}{\left< n_0 \right>} \; + \; \frac{1+f}{p \cdot \left< n_0 \right>}
     \; + \; \frac{\sigma_{\delta_q}^2}{\left< q \right>^2} \cdot 
             \frac{1}{p^2 \cdot \left< n_0 \right>^2}
\end{eqnarray}
where $F$ is the Fano factor ($\equiv \sigma_{n_0}^2 / \left<n_0 \right>$) 
and $f$ is the relative variance of gas gain for single drift electrons
($\equiv \sigma_q^2 / \left< q \right>^2$).
 
Usually the contribution of the third term is negligibly small as far as the
average transmission rate ($p$) is not too small.
In our case, the noise contribution (the width of the pedestal distribution) can be
quadratically subtracted from the signal width.
Therefore, the values of $F$ and $f$ are obtained respectively from the
$y$-intercept and the slope of the straight line fitted through the
resolution squared as a function of $1/p$,
once the value of $\left< n_0 \right>$ is known.
An example is shown in Fig.~A.1 along with a fitted straight line.

%%%%%%%%%%%%%%%%%%%%%%%%%%%%%%%%%%%%%%%%%%%%%%%%%%%%%%%%%%%%%%%%%%%%%%%%%%%%%%
\begin{figure}[htbp]
\begin{center}
\includegraphics[width=14cm,clip]{\figdir/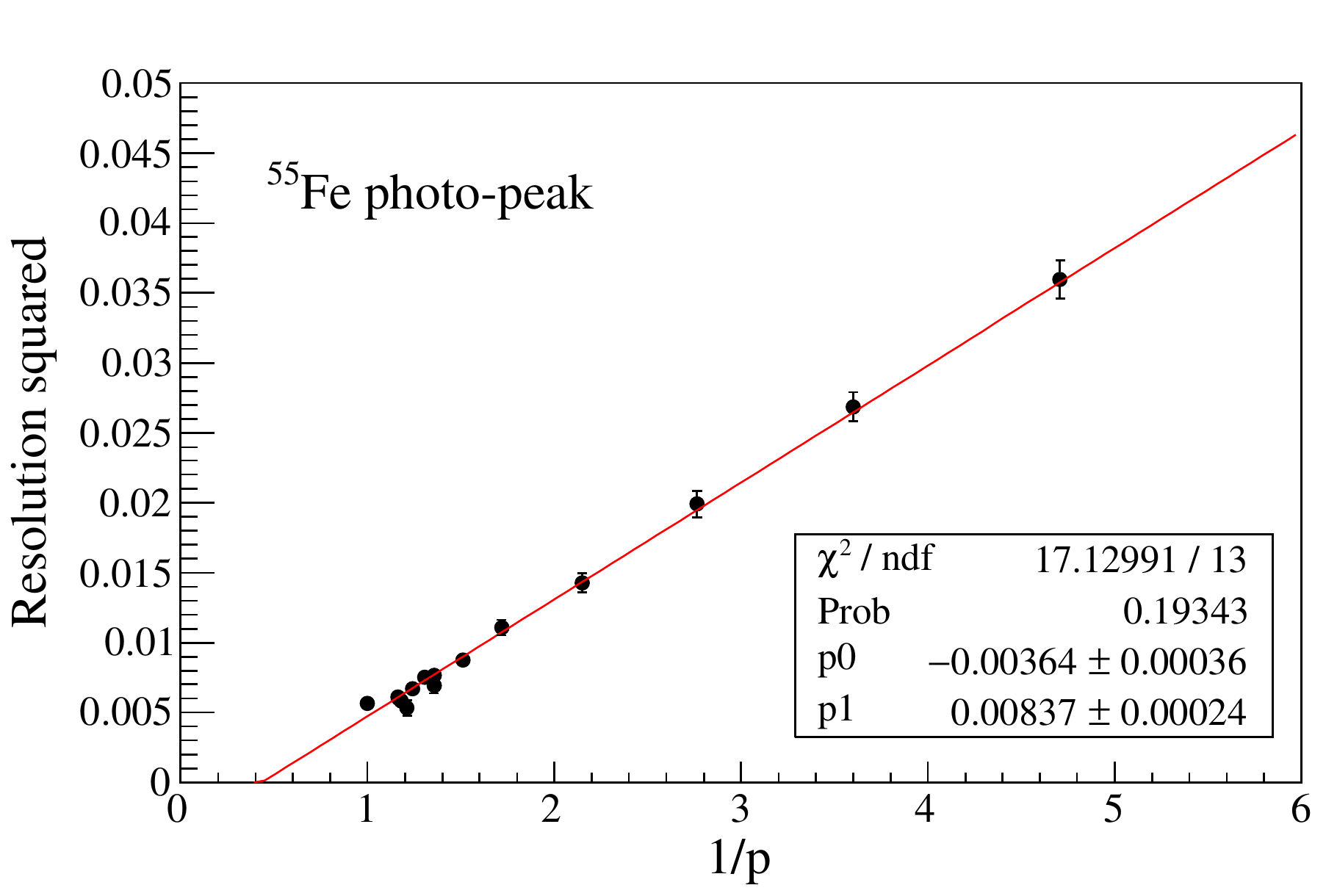}
\end{center}
\vspace{-5mm}
\caption{\label{figa1}
\footnotesize Resolution squared as a function of $1/p$
for the $^{55}$Fe photo-peak
for events with the signal charge deposited exclusively on the
central connector.  
Also shown is a straight line fitted through the data points,
along with the fitted parameters p$_0$ ($y$-intercept) and 
p$_1$ (slope).
The data point for $1/p$ = 1 (ungated signals) was obtained 
with the reversed drift field.
}
\end{figure}
%%%%%%%%%%%%%%%%%%%%%%%%%%%%%%%%%%%%%%%%%%%%%%%%%%%%%%%%%%%%%%%%%%%%%%%%%%%%%%

If we assume 220 for $\left< n_0 \right>$\footnote{
In the rough estimation of $\left< n_0 \right>$
the weighted average energy transferred to electrons from  
the Mn K$_\alpha$ and K$_\beta$ photons
was assumed to be 5.83 keV,
and the $W\/$-values of 26.4 eV, 34.3 eV and 23.4 eV
were used respectively for Ar, CF$_4$, and isobutane \cite{ICRU,Reinking},
without consideration for their stopping powers.
}
the estimated values of $F$ and $f$ are
0.20 $\pm$ 0.08  and 0.84 $\pm$ 0.06,
respectively.
It should be noted, however, that the values of $F$ and $f$ quoted above
are preliminary since the average number of primary electrons
($\left< n_0 \right>$) is not exactly known
for the Penning gas mixture used\footnote{
In fact, larger values of $\left< n_0 \right>$ have been observed by
measurements with pixel readout for argon-isobutane mixtures
(see, for example, Refs.~\cite{Paul, Kaminski}).
}.
Possible increase in the value of $\left< n_0 \right>$ makes $F$ ($f$)
smaller (larger). 
In addition, the collection efficiency of the first GEM
($\equiv p^\prime$) is not yet measured precisely.
If $p^\prime$ is less than unity the average transmission rate $p$
in Eq.~(A.15) needs to be replaced with $p \cdot p^\prime$ 
since the {\it selections\/} made by the gating foil and the first GEM
are independent.  
In that case, the value of $f$ above is an overestimate,
whereas that of $F$ is expected to be unaffected.
It should also be noted that the local (hole-to-hole) average gain
variation within a GEM foil is not taken into account in Eq.~(A.15).

This technique to simultaneously measure the values of $F$ and $f$ is
applicable to other gas-amplification devices such as Micromegas
and/or to other gas mixtures.
We plan to conduct a series of precision measurements
using a dedicated small chamber, and with more statistics
particularly at small transmission rates.

The energy resolution for monochromatic photons gets worse significantly
with decreasing transmission rate ($p$).
From Eq.~(A.15) 
the resolution is expressed as 
\begin{equation}
R^2 = \frac{F + f}{\left< n_0 \right>} 
\end{equation}
for the full transmission ($p$ = 1),
whereas for $p = 0.8$, for example, it is given by
\begin{equation}
R^2 = \frac{F + 0.25 + 1.25 \cdot f}{\left< n_0 \right>} 
\end{equation}
when the noise contribution is negligible.
If we assume the values of 0.20 and 0.84, respectively for $F$ and $f$
the resolution degrades by $\sim$ 20\% with the average transmission
rate of 80\%. 

In the case of specific energy loss (d$E$/dx) resolution
for charged particles,
$n_0$ represents the number of drift electrons detected by a pad row. 
Neglecting the last term, and for  $p \lsim 1$, Eq.~(A.13) gives
\begin{equation}
R^2 = \frac{1 + f - p + p \cdot \sigma_{n_0}^2 / \left< n_0 \right>}
{p \cdot \left< n_0 \right>} 
  \sim \frac{\sigma_{n_0}^2} {\left< n_0 \right>^2} 
\end{equation}
since $\sigma_{n_0}^2 / \left< n_0 \right> \gg 1$
(see, for example, Fig.~3 in Ref.~\cite{Kobayashi3}).    
Consequently, the energy loss resolution of a single pad row is almost
insensitive to the transmission rate.

In d$E$/dx measurements with a TPC one usually uses the truncated (or trimmed) mean 
of the charges deposited on many pad rows ($\sim $220 in the ILD-TPC)
or wires, in order to reduce unfavorable contribution from the Landau tail \cite{Allison-Cobb}.
A simple Monte Carlo simulation shows that the energy loss resolution
obtained with a truncated mean method does depend on the transmission
rate, but to a significantly lesser extent as compared to that for monochromatic photons,
and that the gas gain fluctuation ($f\/$) has a small but detectable
contribution to degrade the resolution.

The d$E$/dx resolutions in the absence and presence of the gating foil 
will be studied in a beam test using the large prototype \cite{Ralf}
in order to confirm the expectation above.

\begin{comment}

\newpage

\end{comment}

%\end{document}


\begin{thebibliography}{99}
\bibitem{Nygren1} David R. Nygren,
          Proposal to investigate the feasibility of a novel concept in particle
          detection, LBL internal note, February 1974, \\
 \url{https://inspirehep.net/record/1365360}.
\bibitem{Nygren2} David R.~Nygren, The Time-Projection Chamber - A new 4$\pi$
  detector for charged particles, PEP-144 in the proceedings of the 1974 PEP summer study,\\
     \url{http://lss.fnal.gov/conf/C740805/p58.pdf}.

\bibitem{Nygren3} Dave~Nygren, The Time-Projection Chamber~-~1975, PEP-198
        	in the proceedings of the 1975 PEP summer study,\\
 \url{http://slac.stanford.edu/pubs/slacreports/reports04/slac-r-190.pdf}.  


\bibitem{Sauli1} F.~Sauli,
               Nuclear Instruments and Methods in Physics Research A 386 (1997) 531,\\
 \url{https://doi.org/10.1016/S0168-9002(96)01172-2}.
\bibitem{Giomataris}  Y.~Giomataris, et al.,
               Nuclear Instruments and Methods in Physics Research A 376 (1996) 29,\\
 \url{https://doi.org/10.1016/0168-9002(96)00175-1}.
\bibitem{ILD} T.~Behnke, J.E.~Brau, P.N.~Burrows, J.~Fuster, M.~Peskin, M.~Stanitzki,
	          Y.~Sugimoto, S.~Yamada, H.~Yamamoto (eds.),
                  The International Linear Collider Technical Design Report 
                  - Volume 4: Detectors (2013), arXiv:1306.6329,\\
 \url{https://arxiv.org/ftp/arxiv/papers/1306/1306.6329.pdf}.
\bibitem{ILC} The International Linear Collider Technical Design Report (2013),\\
 \url{http://www.linearcollider.org/ILC/Publications/Technical-Design-Report}.

\bibitem{Ackermann}  K.~Ackermann, et al.,
               Nuclear Instruments and Methods in Physics Research A 623 (2010) 141,\\
 \url{https://doi.org/10.1016/j.nima.2010.02.175}.
\bibitem{Kobayashi1}  Makoto Kobayashi,
               Nuclear Instruments and Methods in Physics Research A 562 (2006) 136,\\
 \url{https://doi.org/10.1016/j.nima.2006.03.001}.

\bibitem{Arogancia}  D.C.~Arogancia, et al.,
               Nuclear Instruments and Methods in Physics Research A 602 (2009) 403,\\
 \url{https://doi.org/10.1016/j.nima.2009.01.014}.
\bibitem{Kobayashi2}  M.~Kobayashi, et al.,
               Nuclear Instruments and Methods in Physics Research A 641 (2011) 37,\\
 \url{https://doi.org/10.1016/j.nima.2011.02.042}.
\bibitem{Kobayashi3}  Makoto Kobayashi,
               Nuclear Instruments and Methods in Physics Research A 729 (2013) 273,\\
 \url{https://dx.doi.org/10.1016/j.nima.2013.07.028}.

\bibitem{Kobayashi4}  M.~Kobayashi, et al.,
               Nuclear Instruments and Methods in Physics Research A 767 (2014) 439,\\
 \url{https://doi.org/10.1016/j.nima.2014.08.027}.
\bibitem{Yonamine} R.~Yonamine, et al.,
	            Journal of Instrumentation 9 (2014) C03002,\\
 \url{https://doi.org/10.1088/1748-0221/9/03/C03002}.

\bibitem{Madhu}  M.~Dixit, et al.,
               Nuclear Instruments and Methods in Physics Research A 581 (2007) 254,\\
 \url{https://doi.org/10.1016/j.nima.2007.07.099}.
\bibitem{Ralf}  D.~Atti\'e, et al.,
               Nuclear Instruments and Methods in Physics Research A 856 (2017) 109,\\
 \url{https://doi.org/10.1016/j.nima.2016.11.002}.

\bibitem{ILC-Machine} Chris Adolphsen, et al. (eds.),
               The International Linear Collider Technical Design Report
               - Volume 3: Accelerator Part~II: Baseline Design (2013),
               arXiv:1306.6328,\\
 \url{https://arxiv.org/ftp/arxiv/papers/1306/1306.6328.pdf}.

\bibitem{Vogel} A. Vogel, Beam-induced backgrounds in detectors at the ILC, \\
                PhD Thesis, Universit\"{a}t Hamburg, 2008,\\
 \url{http://www-library.desy.de/preparch/desy/thesis/desy-thesis-08-036.pdf}.

\bibitem{Yamashita} T.~Yamashita, et al.,
               Nuclear Instruments and Methods in Physics Research A 283 (1989) 709,\\
 \url{https://doi.org/10.1016/0168-9002(89)91445-9}.
\bibitem{Sauli2} F. Sauli, L. Ropelewski, P. Everaerts,
               Nuclear Instruments and Methods in Physics Research A 560 (2006) 269,\\
 \url{http://doi.org/10.1016/j.nima.2005.12.239}.
\bibitem{Philippe} P. Gros, et al., Journal of Instrumentation 8 (2013) C11023,\\
 \url{https://doi.org/10.1088/1748-0221/8/11/C11023}.
\bibitem{Ikematsu} Katsumasa Ikematsu (on behalf of the LCTPC-Japan Collaboration),
	Development of Large-Aperture GEMs as a Gating Device of the ILC-TPC
	for Blocking Positive Ion Feedback,
        IEEE Nuclear Science Symposium Conference Record (2014) N44-7,\\
 \url{https://doi.org/10.1109/NSSMIC.2014.7431237}.

\bibitem{Schade}  Peter Schade, Jochen Kaminski, for the LCTPC Collaboration,\\
               Nuclear Instruments and Methods A 628 (2011) 128,
               and references cited therein,\\
 \url{https://doi.org/10.1016/j.nima.2010.06.300}.
\bibitem{Ligtenberg}  C.~Ligtenberg, et al.,
               Nuclear Instruments and Methods A 908 (2018) 18,
	and references cited therein,\\
 \url{https://doi.org/10.1016/j.nima.2018.08.012}.
\bibitem{Hilke}  H.J.~Hilke,
               Nuclear Instruments and Methods 217 (1983) 189,\\
 \url{https://doi.org/10.1016/0167-5087(83)90130-8}.
\bibitem{Peisert}  Anna Peisert,
               Nuclear Instruments and Methods 217 (1983) 229,\\
 \url{https://doi.org/10.1016/0167-5087(83)90139-4}.

\bibitem{Arai} D.~Arai, et al., 
	Development of Gating Foils To Inhibit Ion Feedback Using FPC Production
	Techniques, Proceedings of the 4th Conference on Micro-Pattern Gaseous Detectors,\\
 \url{https://agenda.infn.it/getFile.py/access?contribId=60\&sessionId=2\&resId=1\&materialId=paper\&confId=8839}.\\
        Also \\
 \url{https://agenda.infn.it/getFile.py/access?contribId=60\&sessionId=2\&resId=0\&materialId=slides\&confId=8839}
	\,for the presentation slides.

\bibitem{Kobayashi5}  M.~Kobayashi, et al.,
        Nuclear Instruments and Methods in Physics Research A 845 (2017) 236,\\
 \url{https://doi.org/10.1016/j.nima.2016.06.073}.
\bibitem{altro}
A.~Oskarsson, K.~Dehmelt, J-P.~Dewulf, X.~Janssen,
 A.~Junique, L.~Jonsson, G.~De~Lentdecker, B.~Lundberg,
 U.~Mjornmark, L.~Musa, E.~Stenlund, G.~Trampitsch,
 R.~Volkenborn, Y.~Yang, L.~Osterman, 
A General Purpose Electronic readout system
for tests of Time Projection Chambers, equipped with different avalanche multiplication
systems, Eudet-Memo 2008-49,\\
 \url{http://www.eudet.org/e26/e28/e615/e830/eudet-memo-2008-49.pdf}.

\bibitem{altro2} V.~Hedberg, et al., Journal of Instrumentation 10 (2015) C01035,\\
 \url{https://doi.org/10.1088/1748-0221/10/01/C01035}.

\bibitem{Tamagawa} T.~Tamagawa, et al.,
        Nuclear Instruments and Methods in Physics Research A 608 (2009) 390,\\
 \url{https://doi.org/10.1016/j.nima.2009.07.014}.
\bibitem{Biagi} S.F.~Biagi,
        Nuclear Instruments and Methods in Physics Research A 421 (1999) 234,\\
\url{https://doi.org/10.1016/S0168-9002(98)01233-9}. 
\bibitem{Garfield} \url{https://garfieldpp.web.cern.ch/garfieldpp/}.

\bibitem{Gmsh} Christophe~Geuzaine and Jean-Fran\c{c}oi~Remacle,
               International Journal for Numerical Methods in Engineering
               79 (11) (2009) 1309,\\
 \url{https://onlinelibrary.wiley.com/doi/epdf/10.1002/nme.2579}.

\bibitem{Elmer} \url{https://www.csc.fi/web/elmer/}.

\bibitem{Huxley} L.G.H.~Huxley, R.W.~Crompton,
	The Diffusion and Drift of Electrons in Gases,
	Wiley, New York (1974) 
	(Chapter 1 and Chapter 3), and references cited therein.

\bibitem{Christophorou} L.G.~Christophorou, Atomic and Molecular Radiation Physics,
         Wiley-Interscience (1971).

\bibitem{Schultz} G.~Schultz, G.~Charpak, F.~Sauli, Mobilities of positive ions
  in some gas mixtures used in proportional and drift chambers,\\
  Revue de Physique Appliqu\'{e}e (Paris), 1977, 12 (1), pp.67-70,\\
%
 \url{https://doi.org/10.1051/rphysap:0197700120106700}, also \\ 
%
 \url{https://hal.archives-ouvertes.fr/jpa-00244121/document}.

\bibitem{Langevin} P.~Langevin, Sur la th\'eorie du mouvement brownien,\\ 
         Les Comptes Rendus de l'Acad\'emie des sciences 146 (1908) 530.\\
  Also, D.S.~Lemons and A.~Gythiel, On the Theory of Brownian Motion, \\
  American Journal of Physics 65 (1997) 1079 for the translated version
	with an introduction,\\
 \url{https://doi.org/10.1119/1.18725}.  

\bibitem{Hoshina}  K.~Hoshina, et al.,
               Nuclear Instruments and Methods in Physics Research A 479 (2002) 278,
	       Section 5.1, and references cited therein,\\ 
 \url{https://doi.org/10.1016/S0168-9002(01)00908-1}.

\bibitem{Kunst} Thomas Kunst, Bernhardt Gotz, Bernhard Schmidt,
               Nuclear Instruments and Methods in Physics Research A 324 (1993) 127,\\
 \url{https://doi.org/10.1016/0168-9002(93)90971-J}.


\bibitem{Bunemann} O.~Bunemann, T.E.~Cranshaw, J.A.~Harvey,
                   Design of Grid Ionization Chambers,\\    
                   Canadian Journal of Research Vol.27, Sec. A (1949) 191,\\
%
 \url{http://www.nrcresearchpress.com/doi/pdf/10.1139/cjr49a-019}.
%
\bibitem{Bevilacqua} R. Bevilacqua, et al.,
               Nuclear Instruments and Methods in Physics Research A 770 (2015) 64,\\
 \url{https://doi.org/10.1016/j.nima.2014.10.003}.

\bibitem{Blum} W.~Blum, W.~Riegler, L.~Rolandi,
          Particle Detection with Drift Chambers, Springer-Verlag (2008), 
          Chapter 9.

\bibitem{Hilke2}  H.J.~Hilke, Time projection chambers,
     Reports on Progress in Physics, 73 (2010) 116201, Section 3.2.3,\\
 \url{https://doi.org/10.1088/0034-4885/73/11/116201}.

\bibitem{Breskin} A.~Breskin, et al., 
	       Nuclear Instruments and Methods 178 (1980) 11,\\
 \url{https://doi.org/10.1016/0029-554X(80)90853-8}.

\bibitem{Nemethy} Peter~N\'{e}methy, Piermaria~J.~Oddone, Nobukazu~Toge, Akira~Ishibashi, 
	       Nuclear Instruments and Methods 212 (1983) 273,\\
\url{https://doi.org/10.1016/0167-5087(83)90702-0}.

\bibitem{Mormann} D.~M\"{o}rmann, A.~Breskin, R.~Chechik, D.~Bloch,
               Nuclear Instruments and Methods in Physics Research A 516 (2004) 315,\\
 \url{https://doi.org/10.1016/j.nima.2003.08.156}.

\bibitem{Fano}  U.~Fano, 
               Physical Review 72 (1947) 26,\\
 \url{https://journals.aps.org/pr/pdf/10.1103/PhysRev.72.26}.

\bibitem{Chefdeville} M.~Chefdeville,
 A proposal to study gas gain fluctuations in Micromegas detectors, 2009,\\
 \url{http://lappweb.in2p3.fr/\~chefdevi/Work\_LAPP/Gain\_fluctuations/proposal\_gain\_fluctuations.pdf}.

\bibitem{ICRU} International Commission on Radiation Units and Measurements,
               Average Energy Required to Produce an Ion Pair,
               ICRU Report No.~31, Washington~DC (1979).

\bibitem{Reinking} G.F.~Reinking, L.G.~Christophorou, S.R.~Hunter,
  Journal of Applied Physics 60 (1986) 499,\\
 \url{https://doi.org/10.1063/1.337792}.

\bibitem{Paul} D.~Atti\'e, et al.,
Study of avalanche fluctuations and energy resolution with an InGrid-TimePix detector,
talk presented at 12th Vienna Conference on Instrumentation (VCI 2010),\\
 \url{https://indico.cern.ch/event/51276/contributions/2034170/attachments/967018/1373280/2_Colas.pdf}.


\bibitem{Kaminski} Jochen Kaminski, et al., 
GridPix detector with Timepix3 ASIC,
talk presented at the 5th International Conference on Micro-Pattern Gas Detectors (MPGD 2017),\\
\url{https://indico.cern.ch/event/581417/contributions/2522462/attachments/1465797/2265982/GridPix\_TP3.pdf}.

\bibitem{Allison-Cobb} W.W.M.~Allison and J.H.~Cobb, 
	Relativistic Charged Particle Identification by Energy Loss,
	Annual Review of Nuclear and Particle Science 30 (1980) 253,
	and references cited therein, \\
\url{https://doi.org/10.1146/annurev.ns.30.120180.001345}.

\end{thebibliography}
\end{document}